# Conclusion: Perspectives on urban theories

## Denise Pumain, Juste Raimbault

At the end of the five years of work in our GeoDiverCity program[1], we brought together a diversity of authors from different disciplines. Each person was invited to present an important question about the theories and models of urbanization. They are representative of a variety of currents in urban research. Rather than repeat here the contents of all chapters, we propose two ways to synthesize the scientific contributions of this book. In a first part we replace them in relation to a few principles that were experimented in our program, and in a second part we situate them with respect to a broader view of international literature on these topics.

The first part of this concluding chapter is a selection of salient points from our evolutionary theory of urban systems that are discussed in several of the chapters. As many of our results were already published (Pumain et al. 2015; Cura et al. 2017; Pumain & Reuillon 2017)[2] it was possible to confirm them, or to bring more different evidence or contradictory views. For each of these lively research questions, we report the convergent opinions that emerged from the topics discussed, as well as the open and even controversial perspectives for future work. The second part reports a quantitative analysis which arrives at another form of synthesis from the bibliographies of the chapters of the book and their networks of citations. This requires the use of methods for constructing and exploring large digital bibliographical data. Each part gives an overview of the current state of urban science, first of all according to its reported results and then according to the articulations between the conceptions of those who make it. We can easily imagine that in the near future, the second method will become an essential prerequisite for realizing the first, provided that further semantic analysis of the content of papers could be made.

In the United States a recent controversy has been fueled, both by people claiming either that theories about the city were not available (Brenner & Schmid 2014), or that these theories had been constructed mostly from empirical cases and inspiration from the Western world (Robinson, 2016) and demonstrated most of times on the cases of large cities rather than "ordinary" cities and towns. We want to show that the arguments put forward by the instigators of this controversy are in contradiction with the knowledge acquired throughout the course of the history of urban science. We illustrate here a spiral conception of the cumulativeness of knowledge, according to which it is important not to neglect existing theories, which in principle already contain results drawn from a large number of empirical observations, even if their conceptions of society may seem partly out of date. It is therefore just as useful and necessary to propose revisions of old theories as to pretend to bring entirely new ones. Actually, the theories and models presented in

1 Funded by ERC Advanced Grant 269826 (PI Denise Pumain)

2 A series of papers in journals should be added to that short list, as well as PhD Dissertations by Solène Baffi (2016); Clémentine Cottineau (2014); Olivier Finance (2016); Antonio Cosmo Ignazzi (2015); Sebastien Rey-Coyrehourcq (2015); Clara Schmitt (2014) and Elfie Swerts (2013).

this book have revealed themselves widely compatible and complementary.

## 1 Robust grounds for theories and models in an interdisciplinary urban science

We agree with Scott and Storper (2015) when they claim in a recent paper about urban theories that they can identify "dimensions common to all cities without on the one hand, exaggerating the scope of urban theory, or on the other hand, asserting that every individual city is an irreducible special case' (p.1). Indeed, several points of convergence appear in the chapters, regardless of the authors' disciplinary origin. These points mainly characterize the structuration and transformation of systems of cities and summarize what can be called a common urban dynamics. Although this dynamics is complex, and shares properties with other complex systems, it concerns all systems of cities when constituted in quite extensive territories. It admits slightly different but intelligible modalities according to the particular conditions of the historical development of these systems including rather strong path dependence effects. These historical conditions together with the common dynamics constitute the evolutionary theory of urban systems (Pumain, 2018). Because of the common urban dynamics and its rather strong path dependence, this theory authorizes a certain predictability for investigating the future of urban systems.

### 1.1 Urban growth and the hierarchical structuration of systems of cities

The authors of this book have again found that the distribution of the size of cities is always very dissymmetrical, and it can be described by several types of statistical models, such as Zipf's law or the lognormal distribution, both statistically explained by quasi-stochastic models of urban growth (in which urban growth is proportional to city size). These models allow the comparison between systems of cities located in different regions of the world and at different periods of time. The general relevance of these models reveals a great coherence in the adequacy of the size of the cities to that of the territories in which they are located. The science of cities can therefore rely on this knowledge to make forecasts about the future size and number of cities in the short or medium term, and to look for processes common to the evolution of these distributions, in terms of dynamic growth of cities.

Whatever the country and period of observation, there is a complete imbrication of urban dynamics, economic development, innovation waves and the diversity of urban growth trajectories. As a first approximation, cities belonging to the same territory, thus in principle sharing the same rules of socio-economic and cultural functioning, growing on the long-term at about the same rate, even if their growth rates fluctuate sharply on shorter time intervals. A stochastic model like Gibrat's (1931) is a good description of this process. When one observes not only the quantitative evolution of the population of the cities, but also their qualitative transformations (modification of the production and services, occupations, levels of education and skills, cultural and social practices…), there is also a general propensity for these transformations to occur fairly quickly in all parts of the system (Pumain & Saint-Julien 1979; Paulus 2004). According to an evolutionary theory or urban systems, *there is a generic co-evolution process that structures the hierarchical and functional differentiation of interdependent cities within systems of cities.* In terms of the theory of complex systems, this strong coupling in city trajectories is similar but not equivalent to the co-evolution processes observed in biology (Raimbault 2018; Schamp 2010). However it is well explained by the numerous relational networks that are maintained and continuously renewed between cities that interact through the various types of exchanges of their multiple agents and stakeholders. In such well-connected systems, innovations propagate very often according to a hierarchical process that is reactivated each time a new large wave of novelty appears in the modes of economic production or kinds of social organization. Large cities are adapting first to the changes, which on average reach later the medium and smaller towns. This process corresponds to the variations of the exponents of the scaling laws between size of cities and urban attributes over time (see section 1.4 below). Chapter 5 brings a novel confirmation of such a process, showing that even in the networks generated by last cutting edges information technologies, larger cities as effectively more unequal or concentrated in terms of their social relations, as they become more diverse in terms of the structures of each individual's social networks.

It is now recognised that the urbanization process is self-reinforcing, because it is generated from social interactions of all kind taking place in dense urban milieus. Chapter 12 provides evidence from an unprecedented global vision of the physical growth of cities in relation to their demographic expansion showing that this material expansion and its financial value have grown much faster than population. If economic growth is essential to the progress of urbanization, the latter is making an essential contribution to local and global economic development. This expansion is not only a measure of "agglomeration economies" but also the expression of the multiplying power of networking activities. Actually, each discipline tends to formalize this complex evolutionary process in its own terms. Whether named "agglomeration economies" or "increasing returns to scale" or "accelerating pace of life with city size" (West, 2017) or "multiplicative power of networks" as in Chapter 14, the somehow auto-catalytic incentive and trend to a more or less continuous increase of urban population has generated at the same time increasing inequalities in city sizes. Rather than focusing its theories on explaining the growth of "the" city, geography insists we pay attention to their mutual relationships as well as with their embedding territory. From the observation of the quantitative and qualitative co-evolution of urban trajectories, it appears that urbanization is mainly driven by the exploitation of unequal quality and quantity of resources and costs that may vary widely according to city size. For a long time the process that substitutes and updates new products and services more rapidly in large cities than in small towns occurred within small regions or national territories, whereas since the second half of 20th century it has become a global process exploiting the differentials in resource prices and wage levels according to an "international division of labor" (Aydalot, 1976).

## 1.2 General predictability of urban growth and decline

As a result of the lengthy urbanization process there is a huge diversity of urban trajectories, which however obey two main logics. The largest cities are the ones which repeatedly succeed in adopting successive innovations. In this development, they also extend the spatial scope of their activities. In expanding the networks of their activities, they both help and hamper the development of smaller towns and nearby cities by bypassing their customer networks. This last trend is especially reinforced when the acceleration of transport communication accentuate the historical trend to "space-time contraction" leading to a systematic "spatial reorganization" (Janelle, 1969). Thus in the dynamics of systems of cities, there is a tendency to accentuate hierarchical inequalities which has two origins: from the top, large cities develop on average a little more rapidly at the beginning of each innovation wave, and each time the small towns lose relative importance. Thus, at the same time, there is a strengthening of the urban hierarchy from the top and a "simplification from below". In the long run, this dynamic therefore produces population growth in small towns, which tends to be slower than that of the larger ones, with less diversification of their activities. For a long time, the effects of this trend remained not too visible due to the general population growth, which allowed a wide spread of urban growth throughout the system of cities. However, in recent decades in some developed countries, the demographic decline brings out the phenomenon of "shrinking cities", which alarms local officials, observing the devitalization of these cities and fearing for their future evolution. The theory of urban dynamics, however, ensures that such an evolution is quite predictable. In fact, the decline in population growth does not prevent the concentration dynamic from continuing, increasingly opposing growing metropolises and declining small towns.

Some urban trajectories may deviate from these regularities, depending on certain specializations by economic activities in which they have a comparative advantage based on specific deposits (minerals, energy sources are classics, but nowadays cheap labor pools, research capacities or touristic sites are also selective factors of population concentration). As long as the activity remains prosperous, these cities can grow faster than the entire system, but when the activity enters into recession, it may cause these specialized cities to decline faster than others.

The authors of this book share knowledge that may reassures the actors who receive too often some alarmist messages about the urbanization process. Although as in all complex systems the prediction is in theory impossible, we affirm that there is a certain statistical predictability in city growth and city size over short time periods due to the relatively slow time scale of the dynamics of cities. Urban growth is rarely totally explosive and can be rather well anticipated. This relative predictability does not mean that the intervention of urban actors is vane; proactive adaptive strategies (safe imitation of successful ones in similar contexts, or anticipation and risk to find new development niches) are always necessary, in a pervasive context of emulation (or co-opetition). The common critics about the largest metropolises that are too often described as « monstruopolises » are not accurate. In fact, there is a correspondence between the population size of the cities and the size of population in the territories where they are located. Our comparative analyses have demonstrated the robustness and sustainability of most urban systems, even if the rather large variability in their spatial, political and economic organization demonstrates that none of them can be erected as a norm nor optimum. Against a unified alignment on urban planning for competitiveness, Chapter 7 advocates for the promotion of diversity as a more fundamental incentive for sustainable growth. Indeed, through both processes of hierarchical differentiation and functional specialization the resulting geodiversity of cities is perhaps the most historically secure engine of social change. It is in that sense that cities and systems of cities are resilient and can be considered as a product of collective territorial intelligence of humankind. Big cities have more inequality but more resources for the poorest. The geodiversity of cities also means that it accommodates for the various stages of individual trajectories through migration and relocation within cities. Homogeneity of housing stock, demographics or economic activities on the other hand tend to lock city in short cycles of prosperity.

## 1.3 Historical transitions and ecological constraints

The urbanization that has everywhere transformed housing patterns on the planet by grouping populations in increasingly large agglomerations is a process that seems so far irrepressible and irreversible. This process has undergone historically two major acceleration phases, which can be represented as "transitions" that are the social equivalent of the physical bifurcation processes (or phase transition) (Sanders, 2017). The first of these "urban transition" characterizes the emergence of cities, which occurred in all the regions of the world that experienced the Neolithic Revolution, and were sufficiently dense in population, vast in extent and open in terms of circulation (Bairoch, 1985). Archaeologists are sometimes reluctant to accept the terms of "revolution" or "emergence" about the apparition of cities, because of the slowness and progressivity of the associated social changes that were rooted in rural communities. This evolution lasted a few millennia, with many new arrivals and disappearances of cities, though not exceeding a proportion of about 10% of the total world population in the pre-industrial ages (Bairoch, 1985). A second transition took place with the great industrial revolution of the 19th century, which has enormously increased the size of cities (tens of millions of inhabitants instead of just one million for the largest ones before then), and raised the proportion of urban population to more than 50%, 80% are expected at the end of the 21st century according to the UN. In the first post-Neolithic phase, cities were strongly constrained by the resource limitations of their immediate environment, even though they had already invented a way of multiplying their capacity to take advantage of these resources by creating new artifacts and forms of social organization and by exchanging these innovations between different places. Their high vulnerability to natural disasters and the vagaries of conflict, as well as the weakness of their technical means, are sufficient to explain the slowness of this first urban development. In the second phase, cities grew into networks, earlier in the richest and most technologically advanced territories than in the countries colonized by them. In these dominated countries, urbanization occurred partly spontaneously, partly under the influence of colonizing countries, then more rapidly under the effect of strong demographic growth and subsequent economic growth. It is now in some of these territories as large as China and India that the largest cities in terms of population are the most numerous. However the economic weight of cities in developed countries remains predominant. In less than two centuries, dominant habitat patterns across the planet have shifted from small, relatively uniform and spatially dispersed rural cores to considerably larger concentrations of much higher densities and extremely differentiated sizes. During this second phase, urban development has meshed the entire planet with a variety of communication networks of very different natures. All in all, these new forms of habitat seem to cover only a small part of the globe, the built-up areas and the networks that connect them occupy no more than 15% of the earth surface. But their footprint in terms of mineral and energy resources and amount of soil mobilized is much larger, to the point that in terms of ecological footprint it has been estimated that the equivalent of three to eight planets would be needed to raise the level life of all urban dwellers to that of the richest lot at present.

The urbanization process is therefore entering a third stage in its development, during which the environmental constraint is gaining importance. The novelty is that its expression is no longer local, restricting the development capabilities of this or that city, but global. Fortunately the constraint can now be controlled thanks to the existence in all territories of these solidly constituted urban networks, very coherent in their hierarchical organization and their functional complementarities. Such an organization allows at the same time to circulate top down the new international or national regulations for promoting the environmental transition, and bottom up for collecting and disseminating the multiple initiatives and inventions that emerge locally for its concrete realization. However, the knowledge that we currently have in terms of what is called the "urban metabolism" is very largely insufficient to give clear indications as to the urban planning policies that would be able to drive this transition most effectively. Do we need more compact cities, smaller or bigger? The many measures that are being developed to test scaling laws applied to cities can no doubt help to strengthen recommendations.

## 1.4 Scaling laws have societal grounds

Among the recent impulses given to urban research, three have been very much invested by the community of physicists, who proposed to apply their formalisms to cities. Fractal geometry has proved itself to account, much better than classical density measurements, for the morphology of built-up surfaces, spatial distributions of urban activities and city networks, as was mentioned in Chapter 1. The second impulse was caused by the avalanche of massive data collected by mobile or immobile sensors. It has not really given rise yet to the emergence of new theoretical propositions, still mostly describing, with other models, well-known regularities concerning urban mobility, for example. However, Chapter 6 demonstrates that some new model of commuting based on these data can help predicting $CO_2$ emissions with a rather strong accuracy and therefore can bring useful information to urban planners. This work, as well as the one presented in Chapter 5, clearly show that cellphone connections may be a useful proxy of real social interaction, for which until now effective sources of information were too scarce. The fact remains that current approaches are still "socially blind", meaning that people are not differentiated according to the social group to which they belong.

But the third impulse, which consists in seeking the expression of scaling laws in the urban world, has led to new theoretical propositions (Bettencourt &West 2010). Since the authors tend to present this theory as universal, the discussion on this subject remains open. The authors of this book have rejected the idea of the universality of the exponent values of urban scaling laws, by showing their dependence on the data used, and by even identifying intelligible regularities from their variation over time and space. Thus regarding the linear scaling of income level per capita in British cities, Chapter 3 demonstrates that the global statistical picture actually hides the well-known geographical North-South divide between Northern regions having large cities and lower income due to their heavy past of 19$^{th}$ century manufacturing, while Southern cities have not so many large cities but higher income levels. This would be another confirmation of the strong linkage between urban scaling laws and the economic evolution of cities in urban systems, as developed in the evolutionary theory of urban systems. This also suggests that investigations in scaling should be enlarged to those of geographical self-similarities for a better examination of local processes of interactions into the model. In Chapter 4 the authors arrive to the same conclusion with different motives and methods. They also demonstrate the fruitfulness of using the residuals of a model of scaling laws for detecting anomalies or local specificities in location strategies on investment.

Regarding the universality of the interactional urban theory that sustains the model of scaling laws at the level of its fundamental principles (West, 2017), it is more difficult to conclude. At this stage universality cannot signify that urban processes are of the same nature as processes in the physical or living worlds. Institutional regulations may always change (and hopefully they will do) for solving the problems linked with the most harmful consequences of exacerbated urban growth and the difficult challenges of all kind generated by this process. At least, research and debates about physics-driven scaling laws have allowed for extended communication about the long-term effects of city size that were studied by economists, historians and geographers (Bairoch, 1977) and to provide formalized expressions that are useful for comparisons. This trend in multiplying quantified research could ultimately lead to clearer recommendations for urban public policies, for instance by providing more precise quantified measures, as for smarter cities. Moreover, it may become part of a computational shift in social science that will enable a qualitative shift in the knowledge that could be extracted from simulation models as suggested in Chapters 13 and 14.

## 1.5 Do we need urban different theories for each culture or region of the world?

In this book we are referring to the long term evolution of cities as well as to its many global variations, first to throw light on some current debates about urban theories and second because these two requirements are necessary when theorizing about social facts and processes. Actually, all things being equal in terms of the size of a country's surface and population, the models that have been presented show a variability in the organization of urban hierarchies, which are generally more contrasted in recent settlement countries than in those where urbanization has a longer history. In addition, the authors of this book have shown that the results of these comparisons can be unstable and sometimes even contradictory. Indeed, they depend heavily on the number of cities that have been selected to represent a system, and in the way chosen to define them. The solution to this problem appears in chapters 3, 4, 11 and 12: more stable results are obtained and more robust conclusions can be made to test their robustness when choices appropriate to their goals are made for defining cities and systems of cities (see section 1.6 below).

We have demonstrated the usefulness of complexity theories and models for comparing a large variety of urban systems all over the world in another recent book (Rozenblat *et al.* 2018). It is indeed legitimate to start analyzing any urban evolution by filtering from data the evidence of the common urban dynamics. This does not mean that dynamic models are constructed from the Western point of view only, and historians of the urbanization process have since long established that the temporal delay between urban growth in industrialized countries and emerging ones did not mean that the latter would encounter exactly the same sequence as the former (Bairoch 1985). But the same kind of adaptive processes leading to co-evolution of cities are at work in all systems of cities, as exemplified in Chapter 11 with urban cases as different as the Former Soviet Union and South Africa, the first with a background of planned interdependencies, the second marked by a dual system that extend networks because of so many invisible boundaries of the white colonial cities. It may be that in a highly planned country like China, the national and local policies that develop a system of cities that was organized on the same territorial basis for centuries continue its development by obtaining similar patterns as those observed elsewhere. But Chapters 9 and 10 demonstrate each on their own grounds to what extent this reflexive development relies on a specificity of interactions that challenge the current urban and financial theories in their more detailed expression.

## 1.6 Relevant definition and delineation of cities and systems of cities are an essential component of urban science

It is no coincidence that Michael Batty devotes the first part of his Chapter 1 to the question of limits. Many of the controversies and misunderstandings in the literature about the size of cities would be resolved if the authors had been more attentive to the measurement of their size, so to the seemingly ancillary questions of the definition and the delimitation of the objects under study, as in any scientific investigation (Cottineau 2017). Overall, the authors of this book recognize themselves in a two-level ontology that identifies "cities" in "systems of cities" from two scales of relationships in the space-time of societies, that of daily interactions and that of evolutionary interdependencies in the longer term (Pumain 2017). It is true that with the increasing range and diversity of interactions among cities over the last decades, clearly identifying these two levels has become more difficult and their nesting is no longer so strict as in long-established urban theories (Reynaud 1841; Christaller 1933; Berry 1964; Pred 1976).

Cities are complex objects whose multiplicity of definitions refers to the diversity of the interactions that constitute them. These interactions always have a social component, even if they take into account material constraints in terms of the building materials, energy resources or technical and service networks. Some authors are anxious to appear to further objectify their delimitation, based on satellite images of the built space, by naming the forms thus constructed as "natural cities" (Jia & Jiang 2010) or "city proper" (Rosen & Resnick, 1980), which could be considered as a conceptual oxymoron. In fact, the spatial expansion of urban buildings is always linked to social organizations, which impose more or less severe constraints and regulations on the occupation of space by buildings. Cities are objects whose definition is always linked to the political organization of the social groups that build and inhabit them, sometimes identified in administrative definitions and delimitations. Populations and sometimes activities are often enumerated within such limits, as in census data bases. In some cases, some geographical overlaps occur in these urban political definitions, as in the Chinese censuses, which may overestimate the urban populations concerned, while on the contrary the selection of eligible population groups (owning urban residence permits, or urban hukou) leads to their underestimation. This example of a floating population significant in terms of its proportions in urban population typically challenges the "classical" definitions of cities, and is a subject of research in itself (Swerts 2013 and 2017).

Researchers often prefer building their own delineation rather than using official boundaries whose speed of evolution may be slower than that of the spatial expansion of the cities. The concept of *urban agglomeration* has been invented and sometimes documented by statistical institutions to better observe this reality. But the motorization of transport has often led, in all parts of the world, to disrupt the continuity of the urban limits drawn on the ground by connecting very strongly by daily relations with the historic heart of the cities sometimes more distant places, leading to define *functional urban areas*. Sometimes they form groups of intensely connected cities, called *megalopolises*, such as the one identified in 1957 by Jean Gottmann between Boston and Washington (Gottmann 1957), or as those developing in China, between Nanjing and Shanghai and in the Pearl River region around Guangzhou and Shenzhen. These highly integrated *mega-city-regions* are suggested by some researchers such as Le Nechet (2017) as the latest transition of human settlements. Even if the interactions between cities are spreading more and more at multiple scales, even sometimes connecting a small, very specialized town to the whole world, it is not possible to deduce that the delimitation of cities becomes totally obsolete. The fact that technologies, lifestyles and representations tend to become widely influenced by urban cultures everywhere, as Brenner and Schmid (2014) translate into the idea of a "planetary urbanization" cannot support urban theories that would no longer take into account this meso level of geographical organization that we call the city. In this respect, the systematic explorations of different urban boundaries that are made possible today by algorithms applied to detailed georeferenced databases help to better understand the diversity of spatial configurations of populations and urban activities and to bring nuances to the theories which would be too much generalizing (Cottineau 2017; Cottineau et al., 2017; Cottineau et al. 2018)

There is no question of advocating for a single definition. The important thing is to know how to harmonize those used and to choose those that are best adapted to the problem studied. This recommendation is also valid to identify what can be considered as a "system" of cities. That definition means a strong evolutionary interdependence between member cities, still observed with relevance in the context of the current world states, but increasingly uncertain as globalization weaves long-distance networks between places. It has to be reminded that this uncertainty has existed for a long time in the case of the largest cities, since the scope of urban interactions is generally quite strongly correlated with their size. Chapter 12 in this book in a good example of how a harmonized data base with its carefully segmented analysis is able to provide safe results on which urban theories can develop and be tested. And the way is now open for applying in research a real multi-level concept of cities (Rogov & Rozenblat, 2018).
.

## 1.7 Why it is necessary to maintain a plurality of theories and models

Science is a continuous process of creating and revising theories. Theories are proposed for summarizing large sets of empirical observation within a simpler description that is guided by a coherent interpretation, validated in a more or less wide consensus and experience by many scholars and considered at a given moment as a good candidate for a possible explanation. Regarding urban sciences, the existence of a plurality of theories is necessary even if not always well understood.

The plurality of theories may receive a first justification from the epistemology of social sciences. It appears as reflecting the life of science, where a diversity of possible explanations is provided. When thinking in a "realistic" way, the co-existence of different interpretations of the same facts may be attributed to uncertainties in knowledge and ambiguity in facts that are justifying a diversity of explanations, while thinking as the adepts of philosophical constructivism may link this situation to the diversity of opinions, experience or social positions of researchers in science. But one of the major reasons for admitting a plurality of theories in social sciences is that their objects are usually multi-level and highly complex. Regarding cities, the multi-level character is obvious and usually lead to identify three major levels of inquiry within the apparently continuous scales of urban settlements and networks in space and time.

However, the explanatory factors that have to be considered may be different at each level due to their categorization through "emerging properties". There are not yet integrative theories that can provide satisfying explanation for all possible levels of organization. Indeed, multi-level organization is a characteristic feature of complex systems. It is often admitted that if distinct "levels" can be observed in their organization it is because emerging properties arise when the scale of observation is changing, leading to investigate different kinds of factors and processes.

Another source of plurality of theories is that each discipline in social science did elaborate consistent sets of interpretation of the urban realm according to a specific perspective and the particular processes it usually investigates. Building a meaningful explanation of a particular urban case often consists in borrowing explanatory concepts from sociology, economics, geography, history, urban planning and political economy –each of them embedded in their own pattern of understanding of complex systems- and combining them according to an explanatory hierarchy where their weights may vary. We have thus suggested that the urban complexity may also receive a definition including the number of disciplinary concepts that are required to understand a case study.

Urban theories are not enough developed to provide yet irrefutable recipes to urban planners and developers. But urban knowledge does exist, and should not be neglected in advancing hazardous theories. This reflection is a prerequisite in the current race to develop "smarter cities" that are more respectful of the rights of humanity to live in peace and harmony with their environment.

## 2 A citation network analysis to synthesize urban theories

### 2.1 Method and data

We now turn to a quantitative analysis of the relative positioning of disciplines and approaches discussed above. We propose to use citation network analysis as a proxy to understand the structure of that scientific environment, what captures a single dimension of practices but contains relevant information on endogenous disciplinary structures. We use the method and tools of (Raimbault, 2019a) to construct a citation network from the references cited by chapters of this book. The rationale is to reconstruct from the bottom-up the scientific legacy in which each approach situates itself (a citation is a subjective and positioned asset to provide a basis for further knowledge), what is indeed not fully overlapping with the actual content (e.g. captured by semantics, as (Raimbault et al., 2019) show how the two quantifications are complementary).

The bibliography of each chapter was manually indexed to ensure correct citing references retrieval during the data collection process. Furthermore, for performance purposes, but also to ensure a focus of the network content on urban issues, references clearly out of the scope and which would yield a significant part of the initial network totally unrelated to urban theories (the paper on morphogenesis by (Turing, 1990) is a typical example, being anecdotally cited by papers relating to urban issues, but also massively cited by several branches of biology)[3].

The initial corpus contains N = 402 references, and from it the backward citation network at depth two is reconstructed. This means that all papers citing the initial corpus, and a significant proportion of papers citing these citing papers, are collected. This yields a network with V = 596,318 nodes and E = 1,000,604 links. While for performance of data collection reasons, the network is not full (44% of nodes with positive in-degree have all their entering links), the balance between chapters is good (between 39% and 42% when considering chapter subnetworks separately) so this sampling does not bias the analysis. Regarding the language of papers in the networks, running a language detection algorithm on titles (using the python package polyglot) confirms that most of the corpus is in English (80.9%), the second language being Mandarin (4.2%) followed by Spanish (2.4%), German (2.3%), French (2.0%) and Portuguese (2.0%).

## 2.2 Network analysis

We then keep the largest connected component (covering 99.98% of the network) and work on the higher order core of the network, obtained by removing nodes with degree one until no such node is present anymore in the network. The resulting network is smaller (159,648 nodes and 563,956 links) but expected to contain important information in terms of topological structure. A community detection algorithm (Louvain method at fixed resolution of 1) on the symmetrized network is used to reconstruct endogenous disciplines from the viewpoint of citation practices. We obtain 27 communities which have a directed modularity of 0.71. Their size distribution is particular: 16 of them have a cumulated size of less than 1% and can be ignored in the analysis, while the remaining have a rather low hierarchy (rank-size exponent of -0.68 +- 0.08 with an adjusted r-squared of 0.88). This means that communities are rather balanced, confirming that this book covers a broad range of topics with no topic particularly dominating. The main communities are described in Table 1, with their name given after inspection of highest degree papers, their relative size, and some representative papers (chosen among the ones with the highest degree).

3    Code and results are on the open git repository of the project at https://github.com/JusteRaimbault/Perspectivism/tree/master/Models/QuantEpistemo. The raw dataset of the corpus is available on the dataverse at https://doi.org/10.7910/DVN/QCSAKT.

| Community | Relative size (%) | Representative papers |
|---|---|---|
| Regional science | 18.00 | (Cooke & Morgan, 1999) [935] ; (Porter, 2000) [990] |
| Planning/Governance | 12.48 | (Bulkeley, 2013) [541] ; (Healey, 2006) [737] ; (McCann, 2011) [538] |
| Urban Economics | 12.33 | (Gabaix, 1999) [957] ; (Henderson, 1974) [831] |
| Social geography (health, public space, built environment, mode choice) | 11.54 | (Handy et al., 2002) [666] ; (Gehl, 2011) [722] |
| Complexity / Urban simulation / Geosimulation | 8.94 | (Batty, 2013) [642] ; (Waddell, 2002) [893] ; (Benenson & Torrens, 2004) [535] |
| Pattern Design | 7.7 | (Alexander, 1977) [993] |
| Microdemographics | 6.1 | (Bongaarts, 2002) [370] |
| Mobility | 4.6 | (Cresswell, 2006) [701] |
| Transport Networks | 4.2 | (Rodrigue et al., 2016) [438] |
| Spatial Analysis | 4.0 | (Anselin, 2013) [493] |

**Table 1 List of largest citation communities** (covering more than 90% of the network). The name of each was given after inspection of papers of highest degree within the community. We give for each some representative papers among these (degree in brackets).

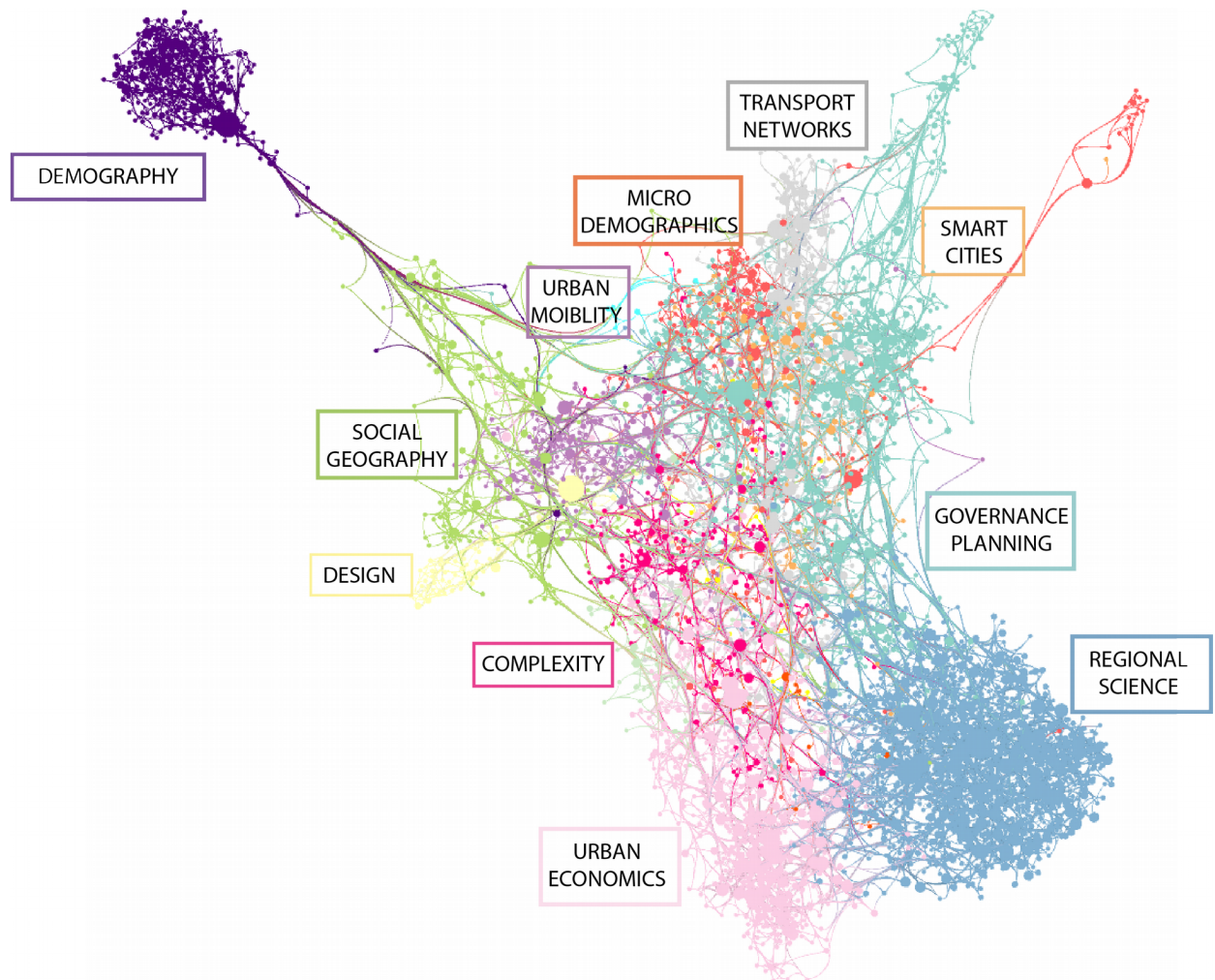

**Figure 1 Visualization of the core citation network.** The network is visualized using the TULIP software, with Fast Multipole Embedder of Martin Gronemann, Curve edges and Edge Bundling algorithms for the layout. In the Fast Multipole Embedder algorithm, attraction forces are mediated in a multilevel way by iterative processing. The color of nodes reflects communities of nodes and their edges. Thanks to Céline Rozenblat who helped in simplifying and improving the readability of the figure.

The content of communities obtained corresponds to some extent to broad disciplinary trends, but also to some thematic structure with some being apparently rather "interdisciplinary". The largest community (called "Regional science") contains works on innovation, firms, clusters, and regions that we attribute to regional science, which is not far but separated from urban economics working on these particular objects and scales. A second community includes work in planning, but also on governance structure and impacts of these (on climate change for example (Bulkeley, 2013). The next cluster is Urban Economics as an expected strongly disciplinary cluster. Then comes works related to social issues, on very different topics (from health to the use of public space, built environment, or transportation mode choice) but all related to the study of the human and social component of the city. An important cluster is then related to complexity and simulation approaches, which can be interpreted more as a "methodological" community. Finally smaller communities can be thematic (Microdemographics, Mobility) or methodological (Spatial analysis). The smallest communities have more chance to being contingent to particular choices or subjects chosen by authors of the book, but the largest components can be seen as a broad overview of urban theory in general. Note that these results remain a partial mapping of urban theories and that many entries remain out of our analysis (urban climate or hydrology for example, population microsimulation models, or purely architectural or urban design approaches, to give a few).

We visualize the network in Fig. 1 in order to have an overview of how the different communities relate to each other. Without surprise, regional science, urban economics, and planning interact strongly and form a very compact triangle. Social geography (in which we can include mobility), and complexity connect also strongly to this core, social geography being mostly connected with planning and complexity, while complexity makes the bridge between urban economics, planning and social geography. Finally, some communities are more isolated at the periphery, such as design or demographics. This visualization confirms that the theories considered in this book are well balanced and relatively well integrated, at least at such a scale of the full citation network. The landscape we get appears broader in its interdisciplinary scope than the graph obtained by Peiris et al. (2018) who analyzed a graph of citations including a smaller number of publications (less than 1500) that were more focused on the topic of systems of cities. Thus, the clusters they identified (regional system, world city network, simulation and complexity, economic geography and city size distribution) are only partially similar to the communities we have found.

The content of largest communities can be studied more precisely, what can also give a better grasp on their level of interdisciplinarity. Therefore, a second community detection can be run within each. The level of modularity then informs whether each community is itself well integrated (low index value) or if it can be decomposed into subfields. As expected, subnetworks are still relatively modular, but with different strengths. Regional science is the least modular community with a modularity of 0.49, urban economics is also relatively low (0.59), while planning (0.63), social geography (0.66) and complexity (0.63) are the most modular communities. This can be interpreted as, for example, urban economics and regional science being more homogeneous in their citation choices. To illustrate how subfields organize, we show in Fig. 2 a visualization of the sub network obtained by keeping the "complexity" community only. We observe a continuum between practical approaches (urban sprawl (Nechyba & Walsh, 2004) and urban growth (Seto et al., 2011) at the bottom), dominating applied simulation approaches (largest communities in the middle, corresponding to models such as Land-use transport interaction models (LUTI) on the left (Waddell, 2002), and cellular automata models on the right (Clarke & Gaydos, 1998)), and more methodological and theoretical approaches at the top, including geosimulation (Benenson & Torrens, 2004), agent-based modeling (Schelling, 1971), urban complexity (Batty, 2007), and urban systems (Pumain, 1996). It is noteworthy to observe the diversity of these "sub-disciplines", but also their complementarity since applied models rely on theoretical and methodological investigations on one side, but also on data-driven investigations on the other side. Furthermore, to connect this diverse community with the rest of the full networks, each sub-community will play its own role in introducing bridges (e.g. applied models will connect to planning, while complexity approaches can connect with economics).

Until now, we performed a mostly visual and descriptive analysis, but it is also possible to quantify the relation between the endogenous disciplines identified, to understand the effective bridges existing or potential integrations. We use for this a basic indicator of inter-citation proportions. Given a total number of citation links made by a given community, we evaluate the proportion of these links made to a paper in another given community. The corresponding matrix for the 5 largest communities is shown in Table 2. The values confirm highly clustered communities, with all having an internal citation rate higher than 77%, the largest being regional science with a rate of 89%. This suggest potential for more bridges (although we quantify here only "direct bridges"; which may miss some intermediate role that would be revealed by centralities e.g. - such an advanced analysis is however out of the scope of this descriptive analysis) between urban theories. One can also distinguish "self-centered" disciplines, in particular regional science, for which the balance of given citation against received citations is always strongly negative, from more open disciplines such as complexity for which it is exactly the contrary. We also confirm the relative positioning discussed with the spatialisation of the network (for example social geography being mostly related to planning and complexity).

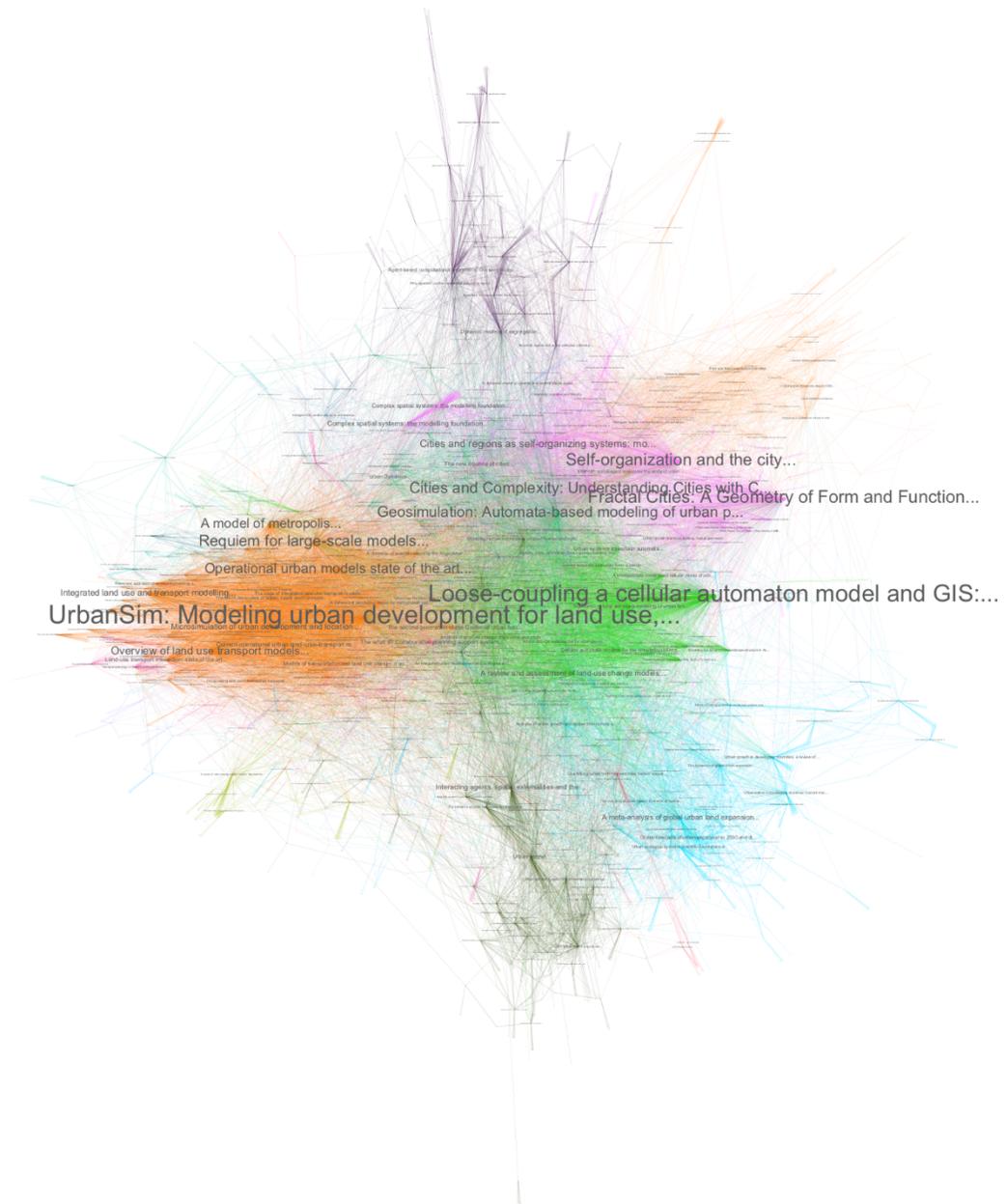

**Figure 2 Visualization of the subnetwork for complexity/simulation models.** Visualization process is the same as before. We observe here communities ranging from data analysis to theoretical and methodological complexity approaches, with applied models in the middle.

|            | Reg. sci. | Planning | Eco. geo. | Social geo. | Complexity | Others |
|------------|-----------|----------|-----------|-------------|------------|--------|
| Reg. sci.  | 89.25     | 2.59     | 4.91      | 0.42        | 0.27       | 2.55   |
| Planning   | 5.15      | 80.32    | 1.96      | 4.35        | 1.71       | 6.51   |
| Urb. Eco.  | 5.77      | 1.90     | 84.12     | 1.08        | 3.18       | 3.96   |
| Social geo.| 1.26      | 5.66     | 2.05      | 78.86       | 4.27       | 10.49  |
| Complexity | 0.88      | 2.69     | 6.87      | 4.79        | 77.38      | 13.19  |

**Table 2 Citation links between main communities.** For the 5 largest communities, proportion (in %) of outcoming citation links in each other community.

Another important insight into the content of this book is then how each chapter is positioned within the network, i.e. how each contributes to the emergence of each different endogenous community. First of all, one can consider subnetworks associated to each chapter. Starting from the references cited by a given chapter, one can reconstruct its subnetwork by getting iteratively citing papers. This produces a subset of the total network as only a subset of the initial corpus was considered. We find that subnetwork sizes range between 113,269 and 139,393 nodes, which corresponds respectively to 71% and 87% of the network, confirming the very high connectivity of branches sprout from different initial seeds. This also confirms a global robustness of the urban theories considered, i.e. that the corresponding scientific practices do refer to a broad common ground. Subnetworks have a high overlap between chapters, as the number of common nodes ranges from 113,133 to 134,467. Focusing on relative overlaps gives some information on the proximity between chapters. The relative overlap is taken as a Jaccard similarity index between sets, that is if N and N' are two sets of nodes, their similarity is given by $J = 2 |N \cap N'| / (|N| + |N'|)$. We show in Figure 3a (above panel) the relative similarity matrix between all chapters. We observe non-intuitive results, as for example (Samaniego, 2019) working on transportation network scaling which relatively does not share much citations with the other chapters on scaling laws and in urban economics. The epistemological chapter dealing with complexities (Raimbault, 2019c) is the farthest from most others, reflecting the difficulty to link meta considerations with applied urban theories. The two chapters on scaling (Arcaute and Hatna, 2019; Finance and Swerts, 2019) intersect mostly between themselves and with the definition of urban complexity (Batty, 2019) and urban economics, but surprisingly not that much with the econophysics chapter (Barthelemy, 2019) which does not refer to a large part of work done on scaling in the field of physics methods applied to urban systems. All in all, we find an absolute high integration, and some unexpected patterns in relative integrations, recalling the contingency of the citation practices that are intrinsic to each scientist with a culture and preferences beside its disciplinary affinities.

Finally, we can study the composition of chapter subnetworks in terms of endogenous communities. Considering a given subnetwork $i$, we compute the probabilities $p_{ij}$ of its nodes to belong to the community $j$. This probability matrix, normalized by taking $p'_{ij} = p_{ij} - <p_{ij}> / std( p_{ij} )$ where average and standard deviation are computed over columns, gives patterns of under or overrepresentation of the different themes within chapters. This normalized matrix is visualized in Figure 3b (bottom panel). We can understand the origin of some communities: for example, demographics mostly come from the chapter on emerging urban systems (Baffi and Cottineau 2019), while a community on settlement data comes from the chapter on urban sprawl (Denis 2019). This also highlights missing entries in some chapters, such as (Raimbault 2019c) which has a very low proportion of urban economics, which is natural given that complexity theories are rather antagonist with the mainstream economics. This also allows finding subtle differences in content, such as the two chapters on scaling, (Finance and Swerts 2019) invoking more spatial analysis in a geography tradition, while (Arcaute and Hatna, 2019) have a relatively higher link to complexity and urban economics. Finally, studying Herfindhal concentration index on composition probability as a measure of "interdisciplinarity" of each chapter does not give significant results (values ranging from 0.84 to 0.86) to differentiate them, and further analysis would be necessary to study this particular aspect (for example using more elaborated indices such as the Rao-Stirling index (Leydesdorff and Rafols, 2011)) but remains out of the scope of this chapter.

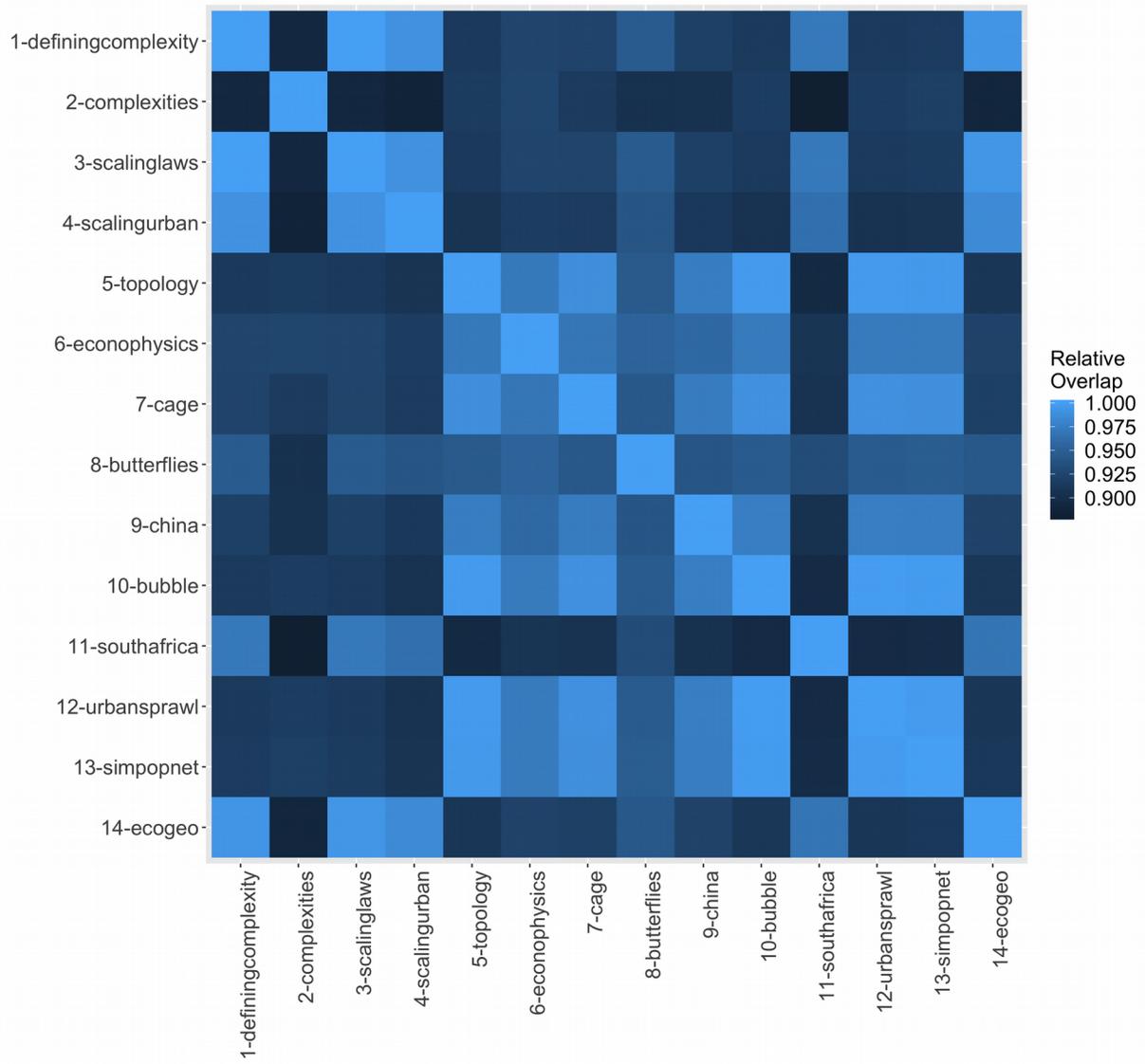

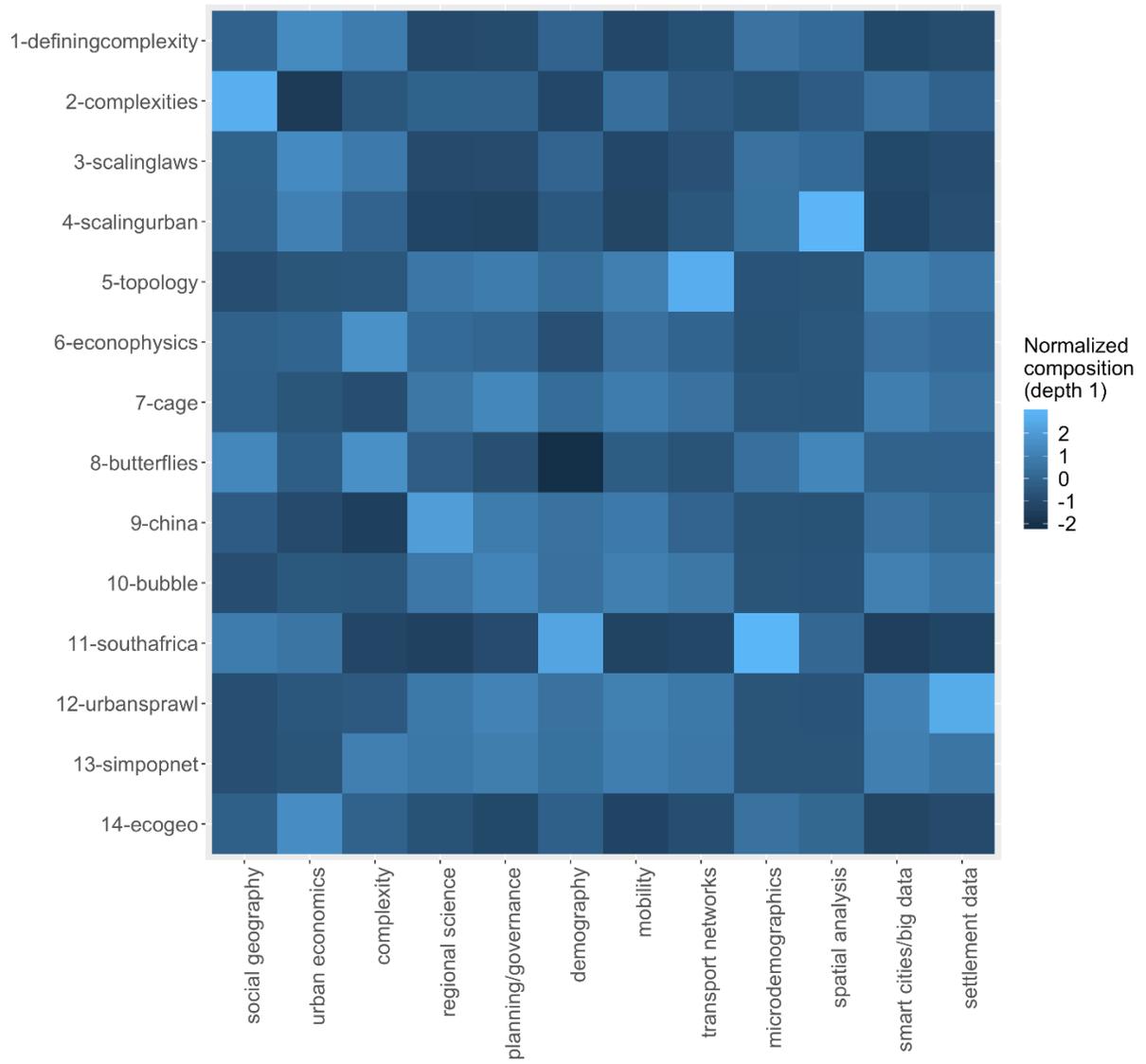

**Figure3 Network coverage and composition of chapters, given by a similarity matrix between chapters (3a), and a composition matrix of chapters in terms of communities (3b).** Proximities between chapters given by a Jaccard similarity index between subnetworks corresponding to each. Chapters are coded the following way: "definingcomplexity" (Batty, 2019); "complexities" (Raimbault, 2019c); "butterflies" (Sanders et al., 2019); "cage" (Bouba-Olga, 2019); "topology" (Samaniego et al., 2019); "econophysics" (Barthelemy, 2019); "scalingurban" (Finance and Swerts, 2019); "scalinglaws" (Arcaute and Hetna, 2019); "china" (Wu, 2019); "southafrica" (Baffi and Cottineau, 2019); "bubble" (Aveline, 2019); "urbansprawl" (Denis, 2019); "ecogeo" (Bida and Rozenblat, 2019); "simpopnet" (Raimbault, 2019d). *(Right)* Composition of chapters in terms of relative share of subnetworks (considering citing papers at the first level only, i.e. papers directly citing the initial corpus) in each community, normalized as center and reduced variables. Negative values correspond to an underrepresentation of the theme while positive values correspond to an overrepresentation.

This analysis allows to better situate each chapter in a global picture of the literature and thus better understand their complementarity. Possible bridges, or new points of view, can also emerge from considering interactions between communities and chapters.

# 3 Modeling and simulation as a medium to couple approaches

In the previous two sections, we gave an overview of how different urban theories can be complementary in theory and in practice. We now discuss why the coupling of heterogeneous approaches is relevant for future urban research and how modeling and simulation could be a powerful medium to do so.

## 3.1 Coupling theories through models

This main proposal is based on general principles for modeling and simulation in the social sciences introduced by Banos (2013), which develops general guidelines to extract knowledge from simulation models. These relate to and draw on widely established practices in diverse disciplines using modeling and simulation, such as ecology (Grimm & Railsback, 2012), computational social science (Epstein, 2006), and general methodological contributions on agent-based modeling for example (Sun et al., 2016). These include in particular that (i) models have different objectives and functions; (ii) they thus must be shared in an open way for their benchmarking and comparison; (iii) models must be reused and coupled; (iv) behavior of models must be known in a precise way with extensive sensitivity analyses. Other principles include for example the need for a strong interaction between models and empirical data, or the fact that problems are most of the time multi-objective and models cannot provide unique optimal solutions, but these have less direct impact on our question. These different aspects are interlinked and form altogether a consistent framework in the spirit of complementary simulation models in an open science and reproducible context. The use of simulation models in itself, beyond all the advantages of being a medium to produce indirect knowledge on processes of a system, is furthermore justified as models are more and more part of the system studied, as Batty (2019) puts it when considering the concept of a "digital twin".

The case of geographical systems, and more particularly urban systems, furthermore justifies the application of these principles, because of their multi-dimensionality, spatio-temporal non-stationarity, multiple aspects of complexity, multi-scalarity. Some aspects of this complexity of urban systems can be specified and linked to Banos' principles. The "ontological complexity" proposed by (Pumain, 2003) as a new alternative to define the complexity of a system, which would be based on the number of viewpoints required to grasp most of system processes, is always high for urban systems, which is equivalent to their high multi-dimensionality. Therefore, the principle of various model objectives and functions is intrinsic to urban systems. The high spatio-temporal non-stationarity (Raimbault, 2019c) and the non-ergodicity (Pumain, 2012) of urban systems directly justify the importance of knowing the model behavior and performing a sensitivity analysis: if the model trajectories are path-dependent or dependent on the application context, an extensive knowledge of model dependency to initial conditions and parameters is essential to extract robust knowledge from it.

Model complementarity and coupling is at the core of (Banos 2013) system of principles. We furthermore argue here that model coupling, in the sense of the construction of integrated models, can be a robust way to couple theories. This can be understood as a sort of "transfer postulate" between theories and associated models. Following (Livet et al. 2010), ontology in the sense of an explicit specification of object and processes studied, is a powerful mediator to build agent-based models of social systems. In this context, different theories would then be mapped to different ontologies, i.e. models, in the modeling domain, and possibly to different methods, tools, data, and empirical analysis. While the latest can be coupled but do not necessarily induce a new knowledge component (coupling two methods is not necessarily a new method, as coupling two empirical analysis does not imply a new one, or it requires generally new models), the coupling of models is particular as elaborating a coupling of models corresponds to constructing a new model: it indeed requires an ontology for coupling processes, even in the case of sequential coupling which is the case where outputs of a first model are used as inputs of a second model (Voinov & Shugart, 2013) . The newly created model should correspond to a new theoretical entity that would then be the coupling of theories. In the epistemological framework of perspectivism proposed by Giere (2010) as an alternative to the opposition between realism and constructivism, this relevance of coupling is furthermore justified: as cognitive agents (scientists) have each their own perspectives, including a purpose, the coupling of perspectives is nothing more than collaborative scientific work. If their disciplinary background strongly differs, such a coupling is a tentative to construct interdisciplinary knowledge. Therefore, our proposal can be linked to "applied perspectivism" described in the third chapter of this book (Raimbault, 2019c).

The construction of such bridges should yield more integrated knowledge, in terms of horizontal integration through the couplings, but also possibly vertical integration if two approaches are at different scales, and furthermore a higher integration of knowledge domains since an increased interaction between them is necessary in the coupling process. Note that a maximum integration is not desirable and would make not much sense, since the practice of deepening knowledge is intrinsically modular and consists in a complex interplay between "disciplinarity and interdisciplinarity" (the virtuous spiral advocated by Banos (2017)). The construction of integrative approaches is thus assumed to participate to a wider context of knowledge production, reinforcing both specific and integrated knowledge. Coupling models only for the sake of it can indeed be counter-productive as pointed by Voinov & Shugart (2013), which differentiate between integration of previously existing models and integration of knowledge from existing models into a new model. This echoes the need to construct a new perspective with its own purpose when coupling two perspectives.

## 3.2 Challenges

We have developed why the coupling of urban theories would be fostered by the coupling of simulation models stemming from these. This would yield integrated approaches, in the sense of a horizontal integration (transversal questions) and a vertical integration (towards multi-scale models) as emphasized by the complex systems roadmap (Chavalarias et al. 2009). We postulate that such approaches are crucial to reach higher standards in evidence-based social sciences, in the sense that they are a path among others to more systematic and evidence-based approaches.

New technical tools and methods will play a crucial role in these integrations. Indeed, as suggested in the previous section, if models are used as intermediaries to couple theories, they however must be well known in terms of behavior, using for example sensitivity analysis methods. In that context, a specific tool and associated methods were developed within the OpenMOLE platform (Reuillon et al. 2013) which provides a workflow engine allowing streamlined model embedding, exploration and distribution of computation on high performance computing environment. These new paradigms and methods are particularly suited to urban issues, as they furthermore arose in the context of the development of urban theories.

We suggest that emerging disciplines in urban science may have a key role to play as integrating approaches. For example, the field of Urban Analytics and City Science coined by (Batty, 2017) when renaming the journal Environment and Planning B Planning and Design, which captures quantitative approaches to urban and territorial systems (with a preferential focus on data analysis methods), is one of these. The new generation of Theoretical and Quantitative Geography inheriting from a long European tradition (Cuyala, 2013) is another branch of these approaches. Geosimulation (Benenson and Torrens, 2004) is also a hybrid and interdisciplinary field which already provided many integrating approaches. The positioning of studies of urban systems by physicists, described as a part of a "physics of society" by Caldarelli et al. (2018) is not clear yet, as they only claim the application of methods from statistical physics to social data and problems, but neither provide directions for such a transfer to be relevant and efficient, nor clarify the elements that would lay the basis for this "new discipline" (for example should they be methodological, with all associated issue of method transfer, or should they be thematic in the sense of object studied, in which case the relation with e.g. urban analytics is not thought).

Many open questions remain, such as the transfer possibilities towards decision making and planning, which can be very different depending on the fields. To what extent confronting approaches can foster the applicability of some is an issue that still has to be investigated. Besides that, it remains impossible to know if some approaches are missed while they could enlighten the particular issues tackled by a candidate integrative approach. The use of quantitative epistemology methods, such as the one used here with citation networks, or multi-dimensional methods (Raimbault et al. 2019), can however help lowering such risks.

## Conclusion

We have in this concluding chapter provided a synthesis of urban theories overviewed through the whole book, by first recalling the most important issues and questions common to most theories of urban systems, which suggested a necessary plurality of such theory. The general explanation of the urbanization of the world is to be found just as much in the capacities of social organization as in economic growth. These two processes are strongly correlated over the long term, even if the process of emergence of innovations who directs the impulses is still difficult to predict, in terms of the conditions of its appearance and its qualitative content. The spatial organization of urban forms is beginning to be better understood, provided we recognize that the explanations and models that account for them are to be conceived as an open dynamics rather than as static equilibrium. The persistence of urban hierarchies, as well as the scaling laws that appear between various attributes of cities, are the product of a dynamic diffusion of innovations that exploits quantitative inequalities and qualitative differences between cities to build complex networks of complementarity and interdependencies, at all levels of city organization and systems of cities. It is in this sense that all people, businesses and local authorities are concerned by the needs of the next important transition for the future of the cities and the urbanization process, which consists of adaptation to climate change and the reduction of fossil resources consumption. More research is needed to construct bridges between theories, in particular the use of simulation models as a powerful medium for interdisciplinary dialogue, as is suggested through the citation network analysis of the scientific landscape around the chapters of the book.

If we may try to convey a specific message from this to the urban citizens, urban planners and stakeholders seeking for general ideas about cities and urbanization, it is rather clear: the current state of urban knowledge results from the collaboration of many disciplines. Urban challenges cannot be solved by a single disciplinary approach or by technologies only. Quantified models can help for solving local problems as well as providing an easier visualization of possible broader scenarios. The new massive observations captured by all kinds of sensors will help for a better local urban management rather than opening entirely new theoretical issues. As cities are fundamentally adaptive complex systems, anchored in a variety of geographical territories and historical contexts, there is not a single norm or model to recommend. On contrary, as in biology, the wide "geodiversity" which is driving urban evolution should be preserved as much as possible in order to maintain an open evolution. Qualitative changes in urbanization cannot be predicted but the future of cities is to be handled with care, precisely because they soon represent our quasi only way of inhabiting our single planet, in a world made more and more interdependent by multiple networks. While the long term effects of accelerated exchanges as in finance and information are not well known, the developing urban interactions could help in sharing solutions rather than exacerbating tensions.